\documentclass[twocolumn]{RAA}
\usepackage{graphicx,times}
\usepackage{natbib}
\usepackage{amssymb,amsmath}
\usepackage{multirow}
\usepackage{booktabs}
\usepackage{threeparttable}
\bibpunct{(}{)}{;}{a}{}{,}

\usepackage[a4paper=true,driverfallback=dvipdfm,pagebackref=true]{hyperref}

\hypersetup{colorlinks = true, linkcolor = green, anchorcolor = red, citecolor = blue, filecolor = red, pagecolor = red, urlcolor = red}

\mathchardef\mhyphen="2D
\defcitealias{Henriques2015}{H15}

\begin{document}
\title{Intrinsic shape variation of quiescent galaxies from redshift 2.5 to 0.5}

\author{Lijun Chen\inst{1,2}, Hong-Xin Zhang\inst{1,2}\thanks{corresponding author}
      }

   \institute{ CAS Key Laboratory for Research in Galaxies and Cosmology, Department of Astronomy, University of Science and Technology of China, Hefei 230026, China; {\it hzhang18@ustc.edu.cn}\\
         \and
             School of Astronomy and Space Sciences, University of Science and Technology of China, Hefei, 230026, China\\
 \vs \no
}

\abstract{
According to the standard ‘inside-out' galaxy formation scenario, galaxies first form a dense core and then gradually assemble their outskirts. This implies that galaxies with similar central stellar mass densities might have evolutionary links. 
We use the UVJ color-color diagram to select quiescent galaxies in the redshift interval from 0.5 to 2.5 and classify them into different subsamples based on their central stellar mass densities, stellar mass, morphological type and redshift.
We then infer the intrinsic axis ratios $\mu_{B/A}$  and $\mu_{C/A}$ of different subsamples based on the apparent axis ratio $q$ distributions, where A, B, and C refers to, respectively, the major, intermediate and minor axis of a triaxial ellipsoidal model. 
We find that 1) massive quiescent galaxies have typical intrinsic shapes similarly close to thick oblate structures, with $\mu_{B/A} \gtrsim 0.9$, regardless of stellar mass, redshift, or central stellar mass densities, and 2) galaxies at higher redshift are systematically thinner than their lower-redshift counterparts, and 3) when splitting the sample into early type and late type with Sersic indices, ETGs at higher redshift are slightly more prolate (smaller average $\mu_{B/A}$) than those at lower redshift. Minor mergers of galaxies may have played important roles in the structural evolution of quiescent galaxies found in this work.
}

   \authorrunning{Chen \& Zhang}            
   \titlerunning{Intrinsic shape of high-$z$ quiescent galaxies}  
   \maketitle

\section{Introduction} \label{sec:intro}
\par
One significant discovery in galaxy astronomy over the past few decades is that high-$z$ galaxies are smaller in size compared to those in the local universe, indicating the size evolution of galaxies across cosmic time (e.g. \citealt{2005ApJ...626..680D, 2007MNRAS.382..109T, 2008ApJ...687L..61B, 2008ApJ...677L...5V, 2011ApJ...743...87W, 2014ApJ...780....1W, 2024ApJ...964..192I, 2024MNRAS.527.6110O}). Local quiescent massive galaxies' half-mass radii are on average 3 times larger than their high-redshift ($z \sim 2$) counterparts (e.g., \citealt{2009ApJ...695..101D, 2019ApJ...880...57M, 2019ApJ...885L..22S, 2019ApJ...877..103S}). Meanwhile, the color gradients of galaxies are rapidly evolving; high-redshift ($z \sim 2$) quiescent galaxies tend to be redder overall, whereas in nearby galaxies ($z \sim 0$) the outskirts are bluer than the centers (e.g., \citealt{2020ApJ...905..170M}). Despite using different data and methodologies, their findings are consistent. Initially, these results sparked some controversy and concerns regarding the quality of photometric redshifts, the depth of imaging data, and the interpretation of broadband spectral energy distributions (SEDs). However, these issues were addressed through deep near-infrared spectroscopy with Gemini/GNIRS and deep imaging data from HST/NICMOS for a sample of massive quiescent galaxies at 
$z \sim 2.3$(\citealt{2006ApJ...649L..71K, 2008ApJ...677L...5V}).

\par
Compared to nearby galaxies, these high-redshift "compact" galaxies stand out because their average central stellar mass densities are approximately 100 times higher than SDSS red galaxies of comparable mass\citep{2008ApJ...677L...5V}. Such massive, high central stellar mass density galaxies are exceedingly rare in the local universe (e.g., \citealt{2009ApJ...692L.118T}), yet in the high-redshift universe($z \sim 2.3$), nearly half of the massive galaxies (\textgreater $10^{11} M_\odot$) are of this type (e.g., \citealt{2006ApJ...649L..71K, 2009ApJ...691.1879W}).

\par
Various scenarios have been proposed to explain the observational phenomena of these high-redshift massive compact galaxies and their subsequent evolution. The most straightforward explanation is that the masses of these galaxies have been overestimated or their sizes underestimated. The mass estimates at the time relied on fitting stellar population templates to the observed spectral or photometric data, and these models carry significant systematic uncertainties. A critical uncertainty lies in the initial mass function (IMF). Previous studies suggested a "bottom-light" IMF (e.g., \citealt{2008MNRAS.385..147D, 2008ApJ...674...29V, 2008MNRAS.385..687W}), which could lead to underestimating the predicted masses, depending on the age of the stellar populations. Another source of uncertainty might come from the estimation of galaxy sizes, which could be underestimated. If there is a strong radial gradient in the mass-to-light ratio of a galaxy, the luminosity-weighted size could differ significantly from the mass-weighted size \citep{2008ApJS..175..356H}. Additionally, limitations in resolution and signal-to-noise ratio might have some impact, albeit minimal.

\par
Another important outcome from observations of high-redshift galaxies is that nearby elliptical galaxies have an average stellar density within the central 1 kpc comparable to that of compact "red and dead" galaxies at high redshift (e.g., \citealt{2009ApJ...697.1290B}). Others using different datasets have reached similar conclusions, noting that the central stellar mass densities or central stellar mass densities of nearby galaxies of equal mass are very close to that of high-redshift compact galaxies (e.g., \citealt{2010ApJ...709.1018V}). In \cite{2010ApJ...709.1018V}, van Dokkum presents beautiful stacked stellar surface density profiles of massive galaxies at different redshifts using NMBS (NEWFIRM Medium Band Survey) data, showed that massive galaxies have gradually built up their outer regions over the past 10 billion years, exhibiting striking uniformity in the stellar mass surface density within a radius of $5kpc$ at different redshifts, while the mass within the $5kpc-75kpc$ range has increased by a factor of about four since $z = 2$. The data suggest that massive galaxies have grown primarily from the inside out, assembling their extended stellar halos around a possibly exponentially profiled compact core. Therefore, the descendants of compact quiescent galaxies at $z \textgreater 2$ may constitute the central parts of today's massive elliptical galaxies.

\par
In hierarchical merger models, mergers are expected to increase stellar mass and/or galaxy size. Major mergers with progenitor mass ratios close to unity result in significant growth in both size and mass, whereas minor mergers involving low-mass companion galaxies are more effective in promoting size growth (e.g., \citealt{2009ApJ...697.1290B, 2009ApJ...699L.178N, 2010ApJ...724..915H}). However, this mechanism requires a very high frequency of minor mergers, many of which involve gas-poor companion galaxies.  Additionally, the colors of high-mass quiescent galaxies are changing dramatically. At z=2, these galaxies are overall red, but by z=0.5, their outskirts are bluer than their centers (e.g., \citealt{2020ApJ...905..170M, 2023ApJ...945..155M}). The primary driver of this outcome is also thought to be minor, gas-poor mergers, which can effectively increase the size of the galaxy and establish color gradients while changing neither the central stellar mass densities nor the stellar mass significantly (e.g., \citealt{2009ApJ...697.1290B, 2009ApJ...699L.178N, 2022ApJ...935..120J}).

\par
In summary, minor mergers are considered a key component in explaining the "inside-out" growth of high-redshift massive galaxies.
Besides, some studies have found that minor mergers are much more common than we previously thought\citep{2012ApJ...746..162N, 2019ApJ...885L..22S, 2020ApJ...899L..26S, 2023ApJ...956L..42S}.

\par
If minor mergers ultimately lead to the evolution of high-$z$ galaxies, then studies on the intrinsic shape can reflect the changes in galaxy morphology caused by minor mergers, thereby shedding light on galaxy formation and evolution. \cite{2012ApJS..203...24V} measured the morphological parameters of a set of high-redshift massive galaxies based on the CANDELS survey and \cite{2014ApJ...792L...6V} studied the intrinsic structure of massive galaxies as a function of redshift and mass. This work also utilizes data from \cite{2014ApJ...792L...6V}, focusing on the intrinsic shape evolution of high-redshift galaxies with evolutionary connections, and attempts to shed light on the evolution of high-redshift galaxies. The rest of the paper is organized as follows. Section \ref{sec:sample} describes the sample selection. Section \ref{sec:infershape} describes the intrinsic shape inference. Sections \ref{sec:result} present our result. Section \ref{sec:summary} summarizes our main conclusion. Throughout this paper, we assume a standard flat {$\Lambda$}CDM cosmology with $H_0$ = 70 $\rm km$ $\rm s^{-1}$ $\rm Mpc^{-1}$ and $\Omega_m$ = 0.3.

\begin{figure*}[htp]
    \centering
    \includegraphics[width=15cm]{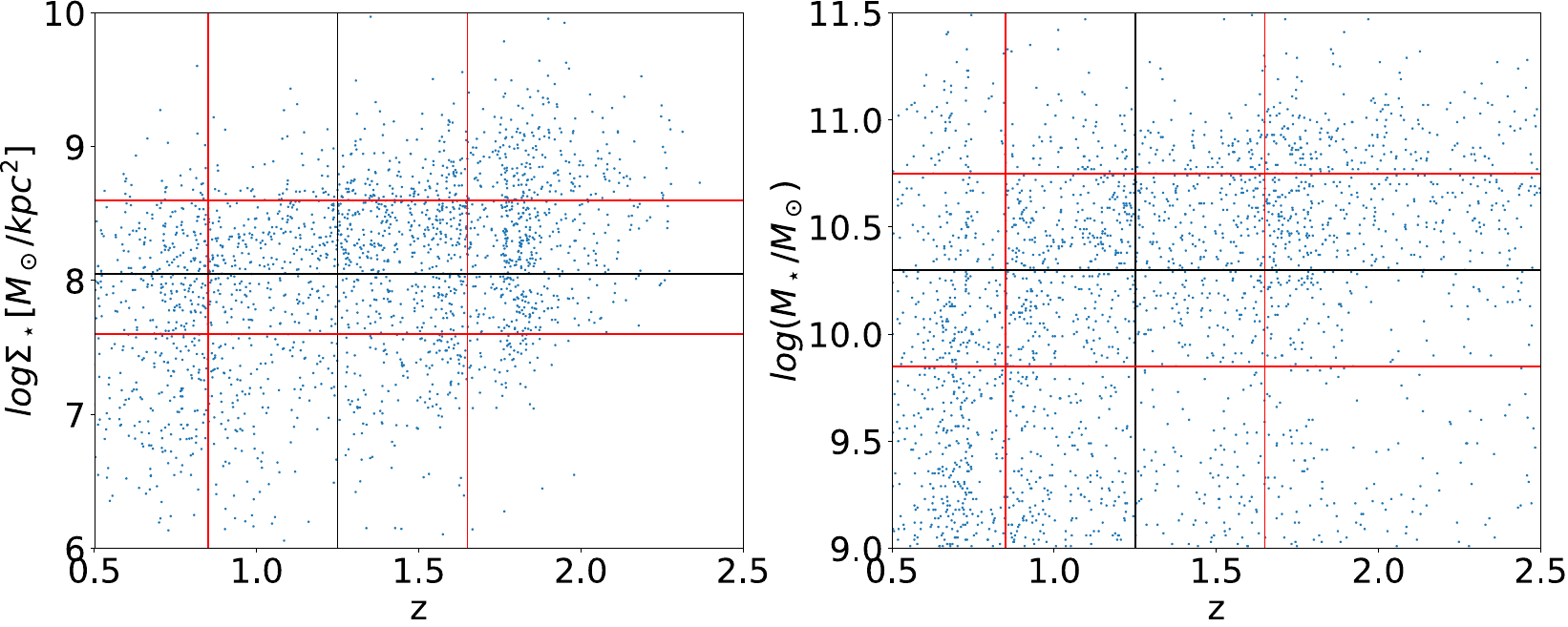}

    \caption{Left: Redshift vs. central stellar mass surface densities of the sample; Right: Redshift vs. stellar mass of the sample. The vertical black lines mark the division between the upper and lower subsamples, and the vertical red lines are the boundaries for the intermediate subsamples.}
    \label{fig:sample}
\end{figure*}

\section{Sample} \label{sec:sample}
  \subsection{Data and Sample}\label{sec:sample selection}

  \par
  The morphological parameters used in this work are derived from \cite{2012ApJS..203...24V}, while the photometric parameters are obtained from \cite{2014ApJS..214...24S}. 
  We use U-V and V-J colors to select quiescent galaxies, as shown in \cite{2009ApJ...691.1879W}.
  A total of 2712 quiescent galaxies with $M_{\star} \textgreater 10^{9} M_{\odot}$ and 0.5 \textless $z$ \textless 2.5 have reliable S\'{e}rsic parameters (i.e. flag value of zero).

  \par
  Utilizing the S\'{e}rsic parameters (i.e., S\'{e}rsic index, total magnitude and half-light radius) from \cite{2014ApJ...792L...6V}, the central surface brightness of galaxies can be derived based on the S\'{e}rsic function. The central surface brightness is then converted into the central stellar mass densities through the overall mass-to-light ratio. The overall mass-to-light ratio is calculated by dividing the total mass by the F\_125W or F\_160W flux. For galaxies with 0.5 \textless $ z $ \textless 2, we use F\_125W based values, for galaxies with 2 \textless $ z $ \textless 2.5, we use F\_160W based values. 
  
  \subsection{Subsample division based on central stellar mass densities and redshift.}
  It is generally challenging to observationally select galaxies with evolutionary connections at different redshifts. There are two commonly adopted methods to select galaxies at different redshifts with possible evolutionary connections, one method invoking the comoving number density of galaxies (e.g., \citealt{2010ApJ...709.1018V}, and the other invoking galaxy central stellar mass densities (e.g., \citealt{2020ApJ...898..171E}). 
  Both of the two methods have limitations. For instance, the first one ignores galaxy mergers, while the second method ignores later gas inflow that may lead to new star formation activities in the centers (likely triggered by wet mergers).
  \par
  As shown in Fig \ref{fig:sample}, we divide the samples into three bins according to redshift, stellar mass, or central stellar mass densities. Specifically, given our limited sample size, we allow for overlap between different bins, The intermediate subsample comprises half of the upper subsample and half of the lower subsample. Although we attempt to use the central stellar mass densities as a probe of galaxies with evolutionary connections at different redshift, the reader should keep in mind the limitation or caveat mentioned above.

\begin{figure*}[htp]
    \centering
    \includegraphics[width=14.5cm]{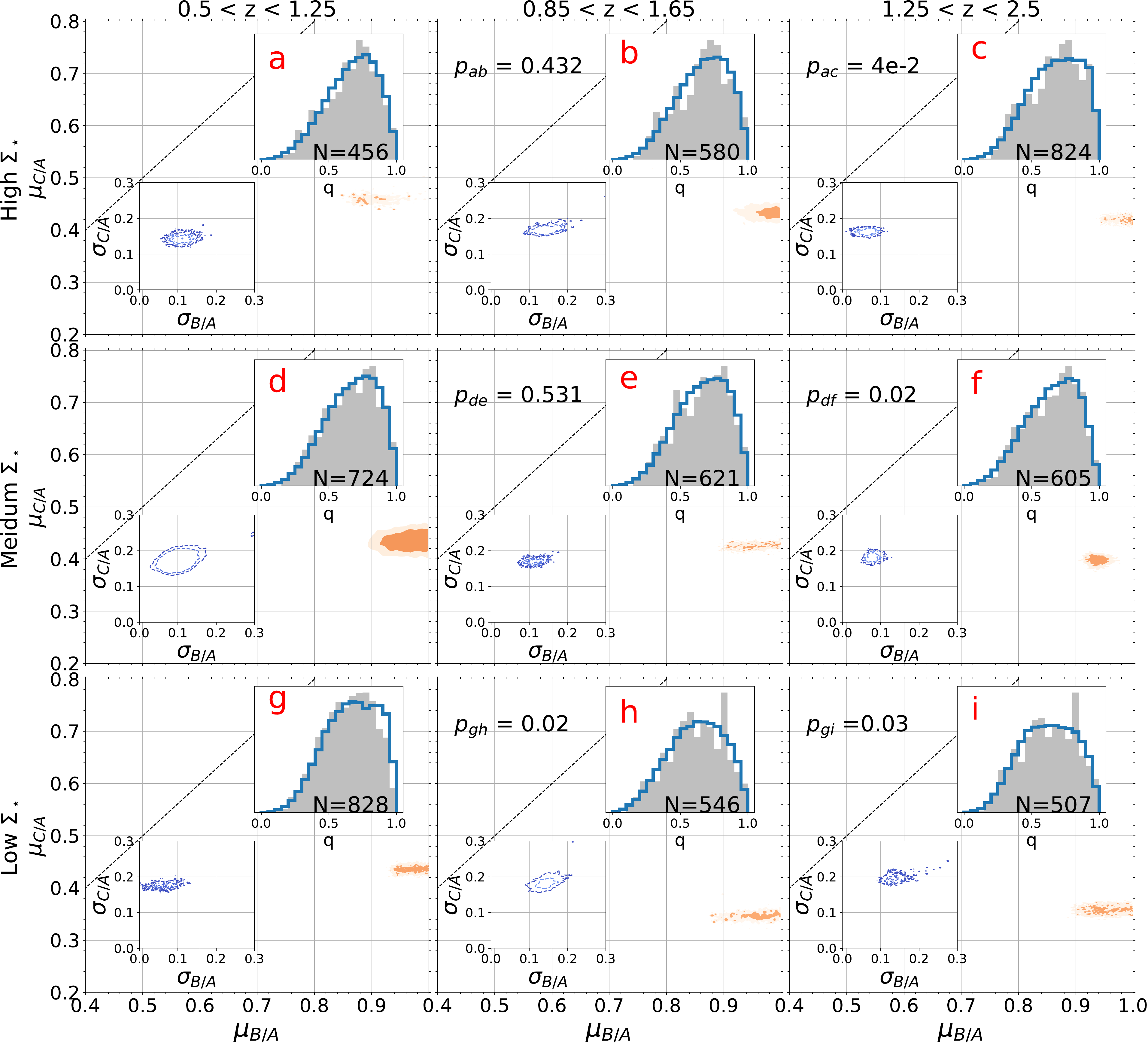}
    \caption{The intrinsic mean axis ratios $\mu_{B/A}$ and $\mu_{C/A}$ inferred for subsamples with different redshift(rows) and central stellar mass densities (columns). In the main plot of each panel, the red and pink filled contours respectively represent the central 68\% and 95\% inter-percentile range of the Bayesian posterior distributions of $\mu_{B/A}$ and $\mu_{C/A}$.\ The black dashed line indicates the B = C line. {The inset plot at the bottom of each panel shows the central 68\%  and 95\% inter-percentile range of the posterior distributions of $\sigma_{B/A}$ (x axis) and $\sigma_{C/A}$ (y axis).} The inset plot at the top of each panel shows the observed axis ratio $q$ distribution (filled grey histogram) and a model distribution obtained by randomly sampling viewing angles of ellipsoids characterized by the most likely mean intrinsic axis ratios and standard deviations (open blue histogram). The number of galaxies in each subsample is indicated at the bottom of each upper-right inset panel. The p-values of the K-S test for the $q$ distributions of the lowest-$z$ (panel ID: a, d, g) and higher-$z$ subsamples in each mass interval are also indicated at the top left corner of each panel.}
    \label{fig:3panel}
\end{figure*}

  \section{Inference of Intrinsic Shape}\label{sec:infershape}
  \par
  Following previous studies, especially \cite{2020ApJ...900..163K} and \cite{2023ApJ...958..117C}, each subsample is modeled using a family of optically-thin triaxial ellipsoids, characterized by their intrinsic axes: major (A), intermediate (B), and minor (C) axes. The intermediate-to-major axis ratio B/A and minor-to-major axis ratio C/A are usually used to represent the ellipsoid shape. We assume that B/A and C/A follow Gaussian distributions for a given sample of galaxies. The mean of B/A $\mu_{B/A}$ and of C/A $\mu_{C/A}$, together with the dispersion of B/A $\sigma_{B/A}$ and of C/A $\sigma_{C/A}$ are the four independent parameters that uniquely quantify the typical intrinsic shape of a sample of galaxies.
  Observed axis ratios, denoted as $q$ (b/a = 1 - ellipticity), represent the projections of the ellipsoids onto the celestial sphere. It is important to note, however, that there exists no direct one-to-one correspondence between $q$ and the intrinsic axis ratios (i.e., B/A and C/A), but rather a statistical relationship. Considering the random projection angle distribution of galaxies belonging to a given population, the relationship between their projected axis ratios ($q$) and their intrinsic shape (i.e., $\mu_{B/A}$, $\sigma_{B/A}$, $\mu_{C/A}$, and $\sigma_{C/A}$) can be readily derived \citep{1998NCimB.113..927S}.

  \par
  Within Bayesian framework, the observed $q$ distribution can be modeled by four parameters: $\mu_{B/A}$, $\sigma_{B/A}$, $\mu_{C/A}$, and $\sigma_{C/A}$, by randomly sampling the orientation angles, where $\mu_{B/A}$ and $\mu_{C/A}$ denote the means of B/A and C/A ratios and $\sigma_{B/A}$ and $\sigma_{C/A}$ the standard deviations of B/A and C/A ratios, respectively. A uniform prior distribution of $\mu_{B/A}$ and $\mu_{C/A}$ from 0 to 1, and $\sigma_{B/A}$ and $\sigma_{C/A}$ from 0 to 0.5 were assumed. Utilizing the $\tt emcee$ package \citep{2013PASP..125..306F}, which implements Markov Chain Monte Carlo (MCMC) sampling methods, we delve into the posterior distributions of the four parameters. A binning resolution of 0.05 is employed for constructing the distributions of $q$ pertaining to both observations and model predictions.

\begin{figure*}[htp]
    \centering
    \includegraphics[width=14.5cm]{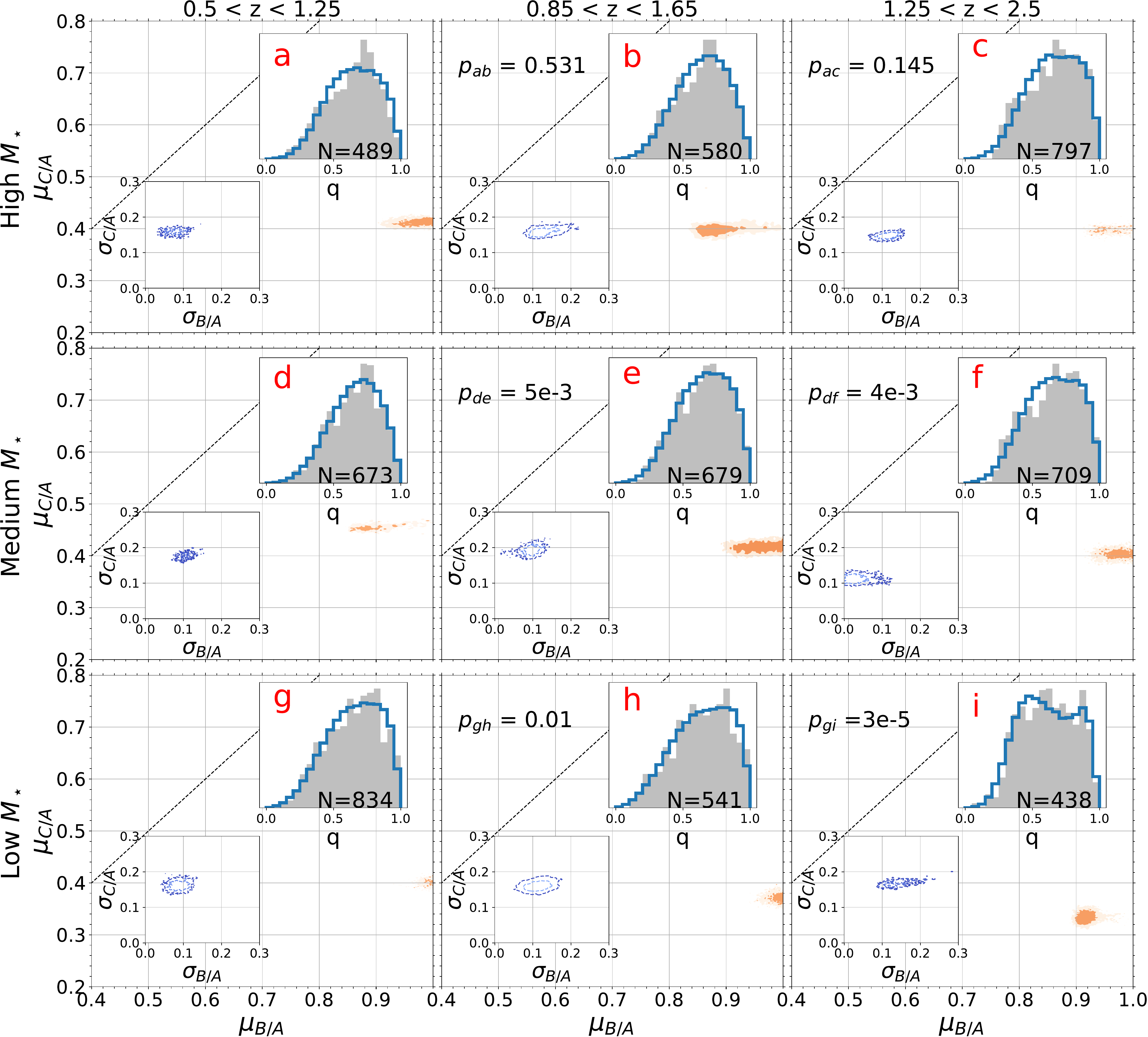}

    \caption{Same as Fig 2 but for subsamples with different redshift (columns) and stellar mass (rows). }
    \label{fig:6panel}
\end{figure*} 

\begin{figure*}[htp]
    \centering
    \includegraphics[width=14.5cm]{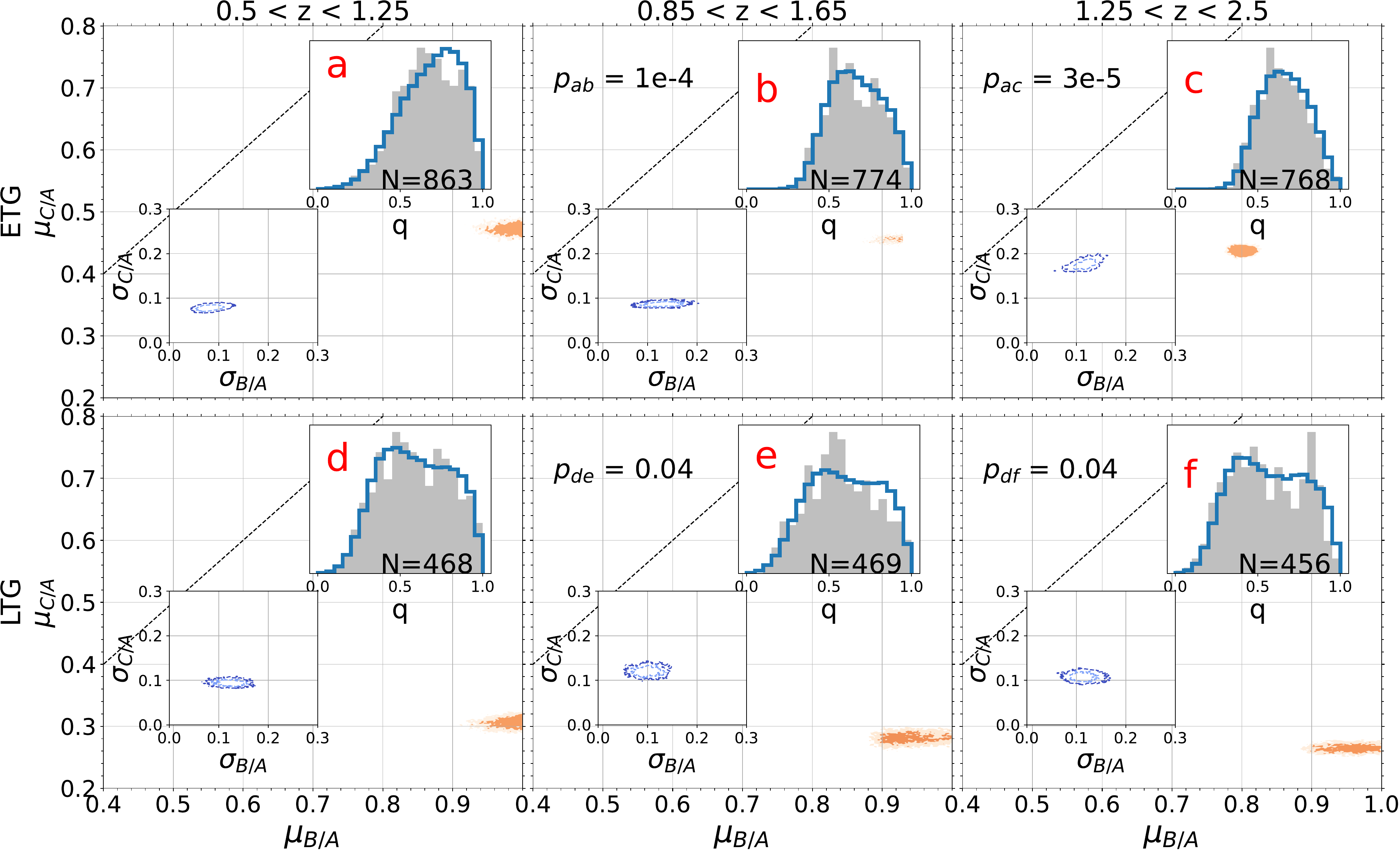}
    \caption{Same as Fig 2 but for ETG, LTG (rows) with different redshifts (columns). }
    \label{fig:ETGLTG}
\end{figure*}

  \section{Results: Intrinsic Shape of High-$z$ galaxies}\label{sec:result}
  
  \par
  First, we considered samples with different central stellar mass surface densities and redshift. The results are shown in Figures \ref{fig:3panel} and Table \ref{table1}. The X and Y axes represent the distribution of $\mu_{B/A}$ and $\mu_{C/A}$, respectively. An inset plot in the bottom left of each panel displays the distribution of $\sigma_{B/A}$ and $\sigma_{C/A}$, and the histogram in the upper right of each panel shows the observed $q$ distribution for each subsample (open) and the inferred posterior $q$ distribution (filled).
  
  \par
  The distribution of $\mu_{B/A}$ and $\mu_{C/A}$ for different subsamples are basically consistent with oblate disky shapes (i.e., $\mu_{B/A}$ $\sim$ 0.9-1.0; $\mu_{C/A}$ $<$ $\mu_{B/A}$).
  It is clear that the average $\mu_{C/A}$ increases from higher to lower redshift (from the right to left panels) for subsamples with different central stellar mass surface densities, indicating that galaxies become thicker at lower redshift. In line with the posterior inference, the peak of the apparent axis ratio $q$ distribution is smaller at higher redshift, suggesting a thinner disk shape at higher redshift. We also quantify the significance of the differences of $q$ distributions by performing K-S test. The K-S test p values are indicated in the figure. The K-S test p values confirm that the q-distributions at different redshift are significantly different, especially when comparing the lowest and highest redshift bins. The insignificant differences between the lowest redshift bins and intermediate redshift bins for the high and medium mass subsamples are probably attributed to the significant overlap in the reshift of the two subsamples.

  \par
  Next, we divide the sample into three subsamples with different stellar mass and explore the stellar mass dependence of intrinsic shape at different redshifts. The results are shown in Figures \ref{fig:6panel} and Tab \ref{table1}, \ref{table2}. Similar to the finding based on subsamples divided by stellar mass surface densities, subsamples at lower redshift tend to have larger average $\mu_{C/A}$. The average $\mu_{B/A}$ does not show monotonic variation with redshift. High mass galaxies at the intermediate redshift, intermediate mass galaxies at the low redshift, and low mass galaxies at the high redshift have significantly smaller $\mu_{B/A}$ ($\sim$ 0.9; still consistent with being an oblate shape) than the other subsamples.

  \par
  Most massive quiescent galaxies are early-type galaxies (ETGs) characterized by concentrated and smooth light profiles, while a smaller fraction is late-type galaxies (LTGs).
  Given the significant morphological differences between ETGs and LTGs, it is helpful to classify and analyze these galaxies separately. We use Sersic indices n $\textgreater$ 2.5 to identify ETGs (e.g., \citealt{2012ApJ...748L..27P}). As showed in Figures \ref{fig:ETGLTG}, ETGs generally are closer to be spherical shape than LTGs, with $\mu_{B/A}$ between 0.8 and 1, and $\mu_{C/A}$ between 0.45 and 0.5. In contrast, LTGs generally exhibit a disk shape with $\mu_{B/A}$ between 0.9 and 1, and $\mu_{C/A}$ between 0.25 and 0.32. The variation with redshift as described in previous sections holds here: low-redshift galaxies are thicker than high-redshift galaxies, regardless of their morphological types. In addition, the $\mu_{B/A}$ decreases as redshift decreases.

    \setlength{\tabcolsep}{1.1em}
    \renewcommand{\arraystretch}{1.1}
    \begin{table*}[]
    \centering
    \caption{$\mu_{B/A}$, $\mu_{B/A}$ and their 16th-84th inter-percentile range for different subsamples analyzed in this work.}
    
    \begin{tabular}{llccccl}
    \hline
    subsample & \multicolumn{2}{c}{0.5 \textless $z$ \textless 1.25} & \multicolumn{2}{c}{0.85 \textless $z$ \textless 1.65} & \multicolumn{2}{c}{1.25 \textless $z$ \textless 2.5} \\
    \cmidrule(l{3pt}r{3pt}){2-3} \cmidrule(l{3pt}r{3pt}){4-5} \cmidrule(l{3pt}r{1pt}){6-7} 
    & {$\mu_{B/A}$} & {$\mu_{C/A}$} & {$\mu_{B/A}$} & {$\mu_{C/A}$} & {$\mu_{B/A}$} & {$\mu_{C/A}$} \\
    \cmidrule(l{3pt}r{3pt}){1-1} \cmidrule(l{3pt}r{3pt}){2-3} \cmidrule(l{3pt}r{3pt}){4-5} \cmidrule(l{3pt}r{1pt}){6-7}
    High $\Sigma_\star$$^{a}$ & $0.893_{-0.025}^{+0.025}$ &
    $0.438_{-0.006}^{+0.006}$ & $0.973_{-0.026}^{+0.026}$ & $0.405_{-0.009}^{+0.009}$ & $0.975_{-0.021}^{+0.021}$ & $0.409_{-0.008}^{+0.008}$ \\
    
    Medium $\Sigma_\star$$^{a}$ & $0.955_{-0.049}^{+0.049}$ & $0.440_{-0.017}^{+0.017}$ & $0.985_{-0.014}^{+0.009}$ & $0.427_{-0.008}^{+0.010}$ & $0.949_{-0.037}^{+0.032}$ & $0.421_{-0.017}^{+0.010}$ \\
    
    Low $\Sigma_\star$$^{a}$ & 
    $0.968_{-0.029}^{+0.029}$ & $0.412_{-0.007}^{+0.007}$ & $0.945_{-0.045}^{+0.045}$ & $0.357_{-0.010}^{+0.010}$ &
    $0.953_{-0.037}^{+0.037}$ & $0.359_{-0.009}^{+0.009}$ \\

    High $M_\star$$^{b}$ & $0.960_{-0.028}^{+0.028}$ & $0.427_{-0.007}^{+0.007}$ & $0.887_{-0.027}^{+0.027}$ & $0.399_{-0.008}^{+0.008}$ & $0.958_{-0.021}^{+0.021}$ & $0.416_{-0.007}^{+0.007}$ \\
    
    Medium $M_\star$$^{b}$ & $0.893_{-0.024}^{+0.024}$ & $0.435_{-0.006}^{+0.006}$ & $0.956_{-0.035}^{+0.035}$ & $0.415_{-0.010}^{+0.010}$ & $0.968_{-0.026}^{+0.026}$ & $0.403_{-0.010}^{+0.010}$ \\
    
    Low $M_\star$$^{b}$ & $0.987_{-0.017}^{+0.017}$ & $0.401_{-0.007}^{+0.007}$ &
    $0.986_{-0.018}^{+0.018}$ & $0.373_{-0.011}^{+0.011}$ & $0.919_{-0.011}^{+0.011}$ & $0.361_{-0.012}^{+0.012}$ \\

    $ETG$$^{c}$ & $0.976_{-0.021}^{+0.017}$ & $0.471_{-0.006}^{+0.007}$ & $0.895_{-0.017}^{+0.037}$ & $0.441_{-0.005}^{+0.005}$ & $0.815_{-0.014}^{+0.015}$ & $0.432_{-0.005}^{+0.007}$ \\
    
    $LTG$$^{c}$ & $0.980_{-0.022}^{+0.013}$ & $0.306_{-0.006}^{+0.006}$ &
    $0.921_{-0.040}^{+0.048}$ & $0.286_{-0.008}^{+0.008}$ & $0.931_{-0.038}^{+0.043}$ & $0.259_{-0.006}^{+0.007}$ \\
    
    \hline
    \end{tabular}

      \begin{tablenotes}
      \small
      \item[a] (a): The result is also plotted in Figure \ref{fig:3panel}; (b): the result is also plotted in Figure \ref{fig:6panel}; (c): the result is also plotted in Figure \ref{fig:ETGLTG}.
      \end{tablenotes}

    \label{table1}
    \end{table*}

    \setlength{\tabcolsep}{1.1em}
    \renewcommand{\arraystretch}{1.1}
    \begin{table*}[]
    \centering
    \caption{$\sigma_{B/A}$, $\sigma_{B/A}$ and their 16th-84th inter-percentile range for different subsamples analyzed in this work.}
    
    \begin{tabular}{llccccl}
    \hline
    subsample & \multicolumn{2}{c}{0.5 \textless $z$ \textless 1.25} & \multicolumn{2}{c}{0.85 \textless $z$ \textless 1.65} & \multicolumn{2}{c}{1.25 \textless $z$ \textless 2.5} \\
    \cmidrule(l{3pt}r{3pt}){2-3} \cmidrule(l{3pt}r{3pt}){4-5} \cmidrule(l{3pt}r{1pt}){6-7} 
    & {$\sigma_{B/A}$} & {$\sigma_{C/A}$} & {$\sigma_{B/A}$} & {$\sigma_{C/A}$} & {$\sigma_{B/A}$} & {$\sigma_{C/A}$} \\
    \cmidrule(l{3pt}r{3pt}){1-1} \cmidrule(l{3pt}r{3pt}){2-3} \cmidrule(l{3pt}r{3pt}){4-5} \cmidrule(l{3pt}r{1pt}){6-7}
    High $\Sigma_\star$$^{a}$ & $0.135_{-0.021}^{+0.021}$ & $0.199_{-0.009}^{+0.009}$ & $0.146_{-0.021}^{+0.021}$ & $0.184_{-0.012}^{+0.012}$ & $0.066_{-0.033}^{+0.033}$ & $0.176_{-0.008}^{+0.008}$  \\
    
    Medium $\Sigma_\star$$^{a}$ & $0.111_{-0.035}^{+0.035}$ &
    $0.177_{-0.019}^{+0.019}$  & $0.111_{-0.035}^{+0.035}$ &
    $0.177_{-0.019}^{+0.019}$ & $0.111_{-0.022}^{+0.022}$ &
    $0.171_{-0.009}^{+0.009}$  \\
    
    Low $\Sigma_\star^{a}$ & $0.097_{-0.019}^{+0.019}$ &
    $0.142_{-0.007}^{+0.007}$ &
    $0.154_{-0.040}^{+0.040}$ &
    $0.173_{-0.013}^{+0.013}$ & $0.113_{-0.022}^{+0.022}$ &
    $0.143_{-0.010}^{+0.010}$ \\

    High $M_\star$$^{b}$ & $0.098_{-0.022}^{+0.022}$ &
    $0.141_{-0.007}^{+0.007}$ & $0.128_{-0.022}^{+0.022}$ &
    $0.161_{-0.009}^{+0.009}$ & $0.078_{-0.022}^{+0.022}$ &
    $0.160_{-0.009}^{+0.009}$  \\
    
    Medium $M_\star$$^{b}$ & $0.134_{-0.020}^{+0.020}$ &
    $0.167_{-0.008}^{+0.008}$ & $0.118_{-0.024}^{+0.024}$ &
    $0.162_{-0.011}^{+0.011}$ & $0.089_{-0.020}^{+0.020}$ &
    $0.163_{-0.011}^{+0.011}$ \\
    
    Low $M_\star^{b}$ & $0.109_{-0.013}^{+0.013}$ &
    $0.177_{-0.009}^{+0.009}$ & $0.098_{-0.022}^{+0.022}$ &
    $0.193_{-0.012}^{+0.012}$ & $0.031_{-0.018}^{+0.018}$ &
    $0.141_{-0.012}^{+0.012}$\\

    $ETG$$^{c}$ & 
    $0.095_{-0.017}^{+0.002}$ &
    $0.082_{-0.005}^{+0.005}$ & $0.082_{-0.020}^{+0.033}$ &
    $0.076_{-0.004}^{+0.004}$ &  
    $0.116_{-0.020}^{+0.019}$ &
    $0.175_{-0.008}^{+0.009}$ \\
    
    $LTG^{c}$ & 
    $0.117_{-0.016}^{+0.015}$ &
    $0.108_{-0.006}^{+0.006}$ &
    $0.130_{-0.028}^{+0.032}$ &
    $0.116_{-0.008}^{+0.009}$ &
    $0.139_{-0.029}^{+0.028}$ &
    $0.097_{-0.007}^{+0.007}$  \\
    \hline
    \end{tabular}

      \begin{tablenotes}
      \small
      \item[a] (a): The result is also plotted in Figure \ref{fig:3panel}; (b): the result is also plotted in Figure \ref{fig:6panel}; (c): the result is also plotted in Figure \ref{fig:ETGLTG}.
      \end{tablenotes}
    \label{table2}
    \end{table*}

\section{Conclusions and Disscusion} \label{sec:summary}
  \par We have selected a sample of massive quiescent galaxies at $0.5 < z < 2.5$ from the CANDELS survey, derived the intrinsic axis ratios for subsamples with different stellar mass and central stellar mass surface densities, and explored the redshift evolution of the intrinsic shapes of different subsamples. We found that the intrinsic shape of these quiescent galaxies is generally consistent with being thick oblate ($\mu_{B/A} \gtrsim 0.9$), galaxies at lower redshift tend to be thicker (i.e., larger $\mu_{C/A}$), for given stellar mass or central stellar mass surface densities. 
  We use S\'{e}rsic indices n = 2.5 to distinguish between ETGs and LTGs. ETGs generally maintain a close-to spherical shape, while LTGs generally maintain a disk shape. ETGs at higher redshift are slightly more prolate (smaller average $\mu_{B/A}$) than those at lower redshift.

  \par
  The structural transformation of galaxies is typically considered to be due to environmental effects or galaxy mergers (e.g., \citealt{2017MNRAS.467.3083R, 2018MNRAS.474.3140M}). Previous studies usually emphasizes the role of major mergers, deeming them an important factor driving changes in galaxy morphology (e.g., \citealt{1977egsp.conf..401T, 2018A&A...614A..66S}). The morphological mix of galaxies exhibits a gradual increase in the fraction of spheroidal systems over cosmic time (e.g., \citealt{1997ApJ...490..577D, 2014MNRAS.444.1125C}). A substantial body of observational evidence indicates that sphericals have recently experienced a major merger (e.g., \citealt{2016MNRAS.463..832W, 2017MNRAS.465.1157R}). 

  \par
  While single major mergers may dramatically change galaxy morphologies, minor mergers are much more frequent and thus may collectively have significant effect in the structural evolution of galaxies (e.g., \citealt{ 2017MNRAS.465.1241W, 2017MNRAS.472L..50M, 2018MNRAS.474.3140M, 2022MNRAS.511..607J}). Many massive spheroids at z $\sim$ 2 lack tidal features that are indicative of major mergers (e.g., \citealt{2014ApJ...780....1W, 2017MNRAS.465.2895L}). It has been shown that multiple minor mergers can redistribute stellar orbits and form spheroidal, slowly rotating galaxies  (e.g., \citealt{2007A&A...476.1179B, 2011A&A...535A...5Q, 2013ApJ...778...61T, 2014MNRAS.444.1475M, 2018MNRAS.473.4956L, 2023A&A...672A..27B}). In extreme cases, even a single minor merger may result in formation of massive spheroids\citep{2019MNRAS.489.4679J}. These observational evidences collectively suggest that minor merger events, and even mini mergers, play important roles in the morphological transformation, especially for massive galaxies.

  \par
  The current study is limited by the range of galaxy stellar mass and coverage of redshift. Galaxies toward the low mass end typically have relatively low surface brightness, which makes the detection challenging in general. To extend the study to galaxies at higher redshift, deep high-resolution imaging surveys at longer wavelength than offerred by the HST surveys would be necessary. To overcome the above two limitations, the imaging surveys that are being performed by James Webb Space Telescope (JWST) will be invaluable and may revolutionize the study of structural properties of galaxies across cosmic time (i.e., \citealt{2023ApJ...956L..42S, 2023ApJ...955L..18L, 2024arXiv241001874N} )

\begin{acknowledgements}

We acknowledge support from the National Key Research and Development Program of China (grant No.2023YFA1608100), and from the NSFC (grant Nos.12122303, 11973039). This work is also supported by the China Manned Space Project (grant Nos. CMSCSST-2021-B02 and CMS-CSST-2021-A07). We also acknowledge support from the CAS Pioneer Hundred Talents Program, and the Cyrus Chun Ying Tang Foundations.

\end{acknowledgements}
  
\bibliographystyle{raa}
\bibliography{raa.bib}


\end{document}